\documentclass[aps,prl,showpacs,twocolumn,%
superscriptaddress,floatfix,preprintnumbers]{revtex4}

\usepackage{epsfig}
\usepackage{amssymb}

\newcommand{\eg}{\textit{e.g.}}

\newcommand{\etal}{\textit{et al.}}
\newcommand{\Lepto}{\textsc{lepto}}
\newcommand{\Pythia}{\textsc{pythia}}

\newcommand{\Pompyt}{\textsc{pompyt}}

\begin{document}


\title{Diffractive Higgs boson production at the
Tevatron and LHC}

\preprint{TSL/ISV-2002-0263}

\author{R.~Enberg}
\affiliation{High Energy Physics, Uppsala University, 
Box 535, S-75121 Uppsala, Sweden}

\author{G.~Ingelman}
\affiliation{High Energy Physics, Uppsala University, 
Box 535, S-75121 Uppsala, Sweden}
\affiliation{Deutsches Elektronen-Synchrotron DESY, 
Notkestrasse 85, D-22603 Hamburg, Germany}

\author{A.~Kissavos}
\affiliation{High Energy Physics, Uppsala University, 
Box 535, S-75121 Uppsala, Sweden}

\author{N.~T\^\i mneanu}
\affiliation{High Energy Physics, Uppsala University, 
Box 535, S-75121 Uppsala, Sweden}

\begin{abstract}
Improved possibilities to find the Higgs boson in diffractive events, having
less hadronic activity, depend on whether the cross section is large enough.
Based on the soft color interaction models that successfully describe
diffractive hard scattering at HERA and the Tevatron, 
we find that only a few diffractive Higgs events may be produced at the 
Tevatron, but we predict a substantial rate at the LHC. 
\end{abstract}

\pacs{14.80.Bn, 12.38.Lg, 13.85.Rm}
\maketitle


The Higgs boson is predicted as the physical manifestation of the mechanism
giving masses to the fundamental particles in the Standard Model. The discovery
of this missing link is of top priority in particle physics. Based on the
discovery \cite{IS,UA8} of diffractive hard scattering processes \cite{StCroix}
it has been considered whether the Higgs can be more easily observed in
diffractive events at high energy hadron colliders. The lower hadronic activity
in such events with large rapidity gaps should improve the possibilities to
reconstruct the Higgs from its decay products. 

In single diffraction a beam hadron emerges quasi-elastically scattered with a
large fraction of the original momentum and separated by a gap in  polar angle
(pseudo\-rapidity $\eta=-\ln \tan \theta/2$) from a produced hadronic system $X$. Particularly clean events are produced in so-called double pomeron exchange (DPE), where both beam hadrons emerge intact separated by rapidity gaps from a centrally produced $X$-system. An extreme possibility is exclusive Higgs production, $p\bar p \to p\bar p H$, where the central system is just a
Higgs boson that may be reconstructed using a missing mass method \cite{Albrow}.
The crucial question for the usefulness of these diffractive Higgs production processes is whether their cross sections are large enough. 

Predicted cross sections vary by orders of magnitude between
calculations based on different models \cite{diffr-Higgs-papers}. 
Some predictions for the Fermilab Tevatron are
large enough to be of experimental interest, whereas others are not. 
The more limited energy, compared to LHC, implies a stronger kinematical
suppression to produce the heavy Higgs boson in such an $X$-system, having only
a fraction of the overall invariant mass of the collision. 
In this
Letter we improve on this theoretical uncertainty by presenting results on
diffractive Higgs production based on the recently developed soft color
interaction models. In contrast to other models used for estimating the
diffractive Higgs cross section, our models have proven very successful in
reproducing experimental data on diffractive hard scattering processes both
from the DESY $ep$ collider HERA and from $p\bar{p}$ collisions at the
Tevatron \cite{SCI-TEV}. This puts us in a good position to give predictions on
diffractive Higgs production.

The soft color interaction (SCI) model \cite{SCI} and the generalized area law
(GAL) model \cite{GAL} were developed in an attempt to better understand
non-perturbative QCD dynamics and provide a unified description of all final
states. The basic assumption is that soft color exchanges give variations in
the topology of the confining color string-fields which then hadronize into different final states, \eg \ with and without rapidity gaps or leading protons. Also other kinds of experimental results are described in a very economical way with only one new parameter. Particularly noteworthy is the turning of a $c\bar{c}$ pair in a color octet state into a singlet state producing charmonium \cite{charmonium} in good agreement with observed rates.

\begin{figure}[h]
\begin{center}
\epsfig{width= 0.9\columnwidth,file=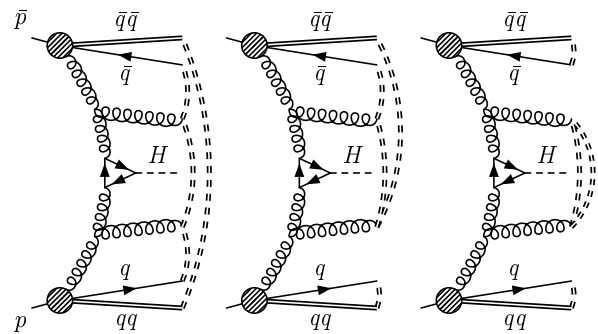,clip=}
\caption{Higgs production in $p\bar{p}$ collisions with string topologies
(double-dashed lines) before and after soft color interactions in the SCI or
GAL model, resulting in events with one or two rapidity gaps 
(leading particles).}
\label{pp-higgs}
\end{center}
\end{figure}


The SCI model \cite{SCI} is implemented in the Lund Monte Carlo programs \Lepto
\ \cite{Lepto} for deep inelastic scattering and \Pythia \ \cite{Pythia} for
hadron-hadron collisions. The hard parton level interactions are given by
standard perturbative matrix elements and parton showers, which are not altered
by the softer non-perturbative effects. The SCI model then applies an explicit
mechanism where color-anticolor (corresponding to non-perturbative gluons) can
be exchanged between the emerging partons and hadron remnants. The probability
for such an exchange cannot be calculated and is therefore taken to be a constant given by a phenomenological parameter $P$. These color exchanges modify the color connections between the partons and thereby the color string-field topology, resulting in different final states after the standard Lund model \cite{lund} has been applied for hadronization (Fig.~\ref{pp-higgs}). 

The GAL model \cite{GAL} is similar in spirit, but is formulated in terms of
interactions between the strings and not the partons. Soft color exchanges
between strings also change the color topology resulting in another string
configuration (Fig.~\ref{pp-higgs}). A generalization of the area law suppression $e^{-bA}$ in the Lund model gives the probability for two strings to interact as $P=P_0[1-\exp{(-b\Delta A)}]$ depending on the resulting change $\Delta A$ of the areas swept out by the strings in momentum space. The exponential factor favors making ``shorter'' strings, \eg, events with gaps, whereas making `longer' strings is suppressed. The fixed probability for soft color exchange in SCI is thus in GAL replaced by a dynamically varying one. 

The Monte Carlo implementations of SCI and GAL
generate complete events with final state particles.
This allows an experimental approach to classify events depending on the final
state: \eg, gaps or no-gaps, leading (anti)protons, charmonium etc. Thus, one
obtains predictive models where a single parameter ($P$ and $P_0$), regulating
the amount of soft color exchanges, has a universal value determined from HERA
rapidity gap data. 

The SCI and GAL models give various diffractive hard scattering processes by
simply choosing different hard scattering subprocesses in \Pythia. Rapidity gap
events containing a $W$, a dijet system or bottom quarks are found to be in
agreement with Tevatron data \cite{SCI-TEV}. The CDF data \cite{CDF-DPE} on dijets in DPE events are also reproduced \cite{SCI-TEV,SCI-DIS2002}, both in cross section and more exclusive quantities such as the dijet mass fraction.
Thus, our models successfully pass these tests given by processes with similar dynamics as diffractive Higgs production. 

The properties of the Higgs boson in the Standard Model are fixed, except for
its mass. The present lower limit is 114.1 GeV and $\chi^2$ fits to electroweak data favors $m_H < 212$ GeV \cite{Higgslimit}. The latest LEP data give an indication ($\sim 2.1\, \sigma$) of a Higgs with a mass of 115.6 GeV \cite{Higgsevidence}. We therefore use $m_H=115$ GeV as our main alternative, but also consider $m_H$ up to 200 GeV. 

Higgs production at
the Tevatron and the LHC can proceed through many subprocesses,
which are included in \Pythia \ version 6 \cite{Pythia}. The dominant one is
$gg\to H$, which accounts for 50\% and 70\% of the cross section (for
$115<m_H<200$ GeV) at the Tevatron and LHC, respectively. In this process, see
Fig.~\ref{pp-higgs}, the gluons couple to a quark loop with dominant
contribution from top due to its large coupling to the Higgs. Other production
channels are 
$q_i\bar{q}_i\to H$, 
$q_i\bar{q}_i\to Z\, H$, 
$q_i\bar{q}_j\to W\, H$, 
$q_iq_j\to q_kq_l H$ and  
$gg\to q_k \bar{q}_k\, H$. 
Their relative contributions depend both on the Higgs mass and the cms
energy. The overall cross sections are obtained by folding the subprocess cross
sections with the parton density distributions (we use CTEQ5L \cite{CTEQ}).
This basic factorization is proven for inclusive hard scattering processes. It
is assumed to also hold in our model since the soft hadronization processes
should not influence the cross section for the hard subprocess, but only affect
the distribution of hadrons in the final state. 
%
\begin{figure}
\begin{center}
\epsfig{width= 0.9\columnwidth,file=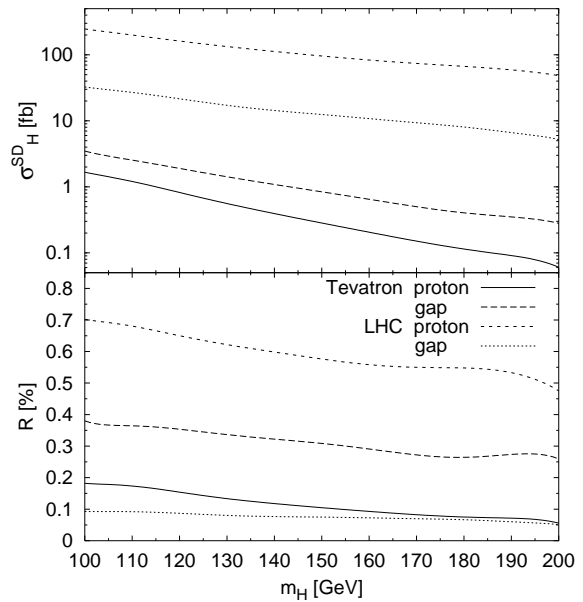}
\caption{Predictions from the SCI model for Higgs production in single
diffractive events defined by a leading proton or rapidity gap criterion at the
Tevatron and LHC. Absolute cross sections and relative ratio (single
diffractive to all Higgs) versus the Higgs mass.}
\label{fig-massdependence}
\end{center}
\end{figure}

After the standard parton showers in \Pythia, SCI or GAL is applied using the
parameter values $P$=0.5 and $P_0$=0.1, respectively. This gives a total sample
of Higgs events, with varying hadronic final states. Single
diffractive (SD) Higgs events are selected using one of two criteria: (1) a
leading (anti)proton with $x_F>0.9$ or (2) a rapidity gap in $2.4<|\eta|<5.9$
as used by the CDF collaboration. Applying the conditions in both hemispheres
results in a sample of DPE Higgs events. The resulting cross sections and
relative rates are shown in Fig.~\ref{fig-massdependence} as a function of the
Higgs mass and for $m_H=115$ GeV in Table \ref{tab-higgs}. The results have an
uncertainty of about a factor two related to details of the hadron remnant
treatment and choice of parton density parameterization.

The cross sections at the Tevatron are quite low in view of the luminosity to
be achieved in Run~II. Higgs in DPE events are far below an observable rate. For
$m_H=115$ GeV, only tens of single diffractive Higgs events are predicted. Only
the most abundant decay channel, $H\to b\bar{b}$, can then be of use and a very
efficient $b$-quark tagging and Higgs reconstruction is required. The
conclusion for the Tevatron is that the advantage of a simplified
reconstruction of the Higgs in the cleaner diffractive events is not really
usable in practice due to a too small number of diffractive Higgs events being
produced.

\begin{table}
\caption{Cross sections at the Tevatron and LHC for Higgs in single diffractive
(SD) and DPE events, using leading proton or rapidity gap definitions, as well
as relative rates (SD/all and DPE/SD) and number (\#) of events, obtained from the
soft color exchange models SCI and GAL.\label{tab-higgs} }
\begin{ruledtabular}
\begin{tabular}{lcccc}
               &  \multicolumn{2}{c}{Tevatron}                       
               &  \multicolumn{2}{c}{LHC} \\
$m_H=115$~GeV  &  \multicolumn{2}{c}{$\sqrt{s}=1.96$~TeV}  
               &  \multicolumn{2}{c}{$\sqrt{s}=14$~TeV} \\
               &  \multicolumn{2}{c}{${\cal L}=20~\mbox{fb}^{-1}$}   
               &  \multicolumn{2}{c}{${\cal L}=30~\mbox{fb}^{-1}$} \\
\hline
$\sigma [\mbox{fb}]$ Higgs-total    & \multicolumn{2}{c}{600}        & \multicolumn{2}{c}{27000} \\
\hline 
\hline
               & SCI & GAL & SCI & GAL \\
\hline 
\multicolumn{5}{l}{Higgs in SD:}\\
$\sigma \; [\mbox{fb}]$ leading-p   & 1.2 & 1.2 & 190 & 160 \\
$\sigma \; [\mbox{fb}]$ gap         & 2.4 & 3.6 & 27 & 27   \\
$R \; [\%]$ leading-p               & 0.2 & 0.2 & 0.7 & 0.6 \\
$R \; [\%]$ gap                     & 0.4 & 0.6 & 0.1 & 0.1 \\
\# H + leading-p                    & 24 & 24 & 5700 & 4800 \\
$\hookrightarrow$ \# H $\rightarrow \gamma\gamma$  & 0.024 & 0.024 & 6 & 5 \\
\hline 
\multicolumn{5}{l}{Higgs in DPE:}\\
$\sigma \; [\mbox{fb}]$ leading-p's & $1.2 \cdot 10^{-4}$   & $2.4 \cdot 10^{-4}$   & 0.19 & 0.16 \\
$\sigma \; [\mbox{fb}]$ gaps        & $2.4 \cdot 10^{-3}$   & $7.2 \cdot 10^{-3}$   & $2.7 \cdot 10^{-4}$   & $5.4 \cdot 10^{-3}$ \\
$R \; [\%]$ leading-p's             & 0.01                  & 0.02                  & 0.1                   & 0.1 \\
$R \; [\%]$ gaps                    & 0.1                   & 0.2                   & 0.001                 & 0.02 \\
\# H + leading-p's                  & 0.0024                & 0.0048                & 6                     & 5 
\end{tabular}
\end{ruledtabular}
\end{table}

\begin{figure}
\begin{center}
\epsfig{width= 1\columnwidth,file=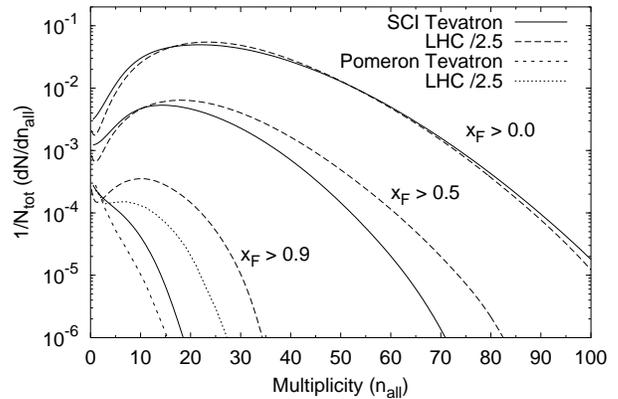}
\caption{Multiplicity (for LHC divided by 2.5) in the region $2.4<|\eta|<5.9$
in the hemisphere of a leading
proton with the indicated minimum $x_F$, for Higgs events from the SCI and the
pomeron models.}
\label{fig-multiplicity}
\end{center}
\end{figure}

In contrast, the high energy and luminosity available at the LHC facilitate a
study of single diffractive Higgs
production, where also the striking $H\to \gamma \gamma$ decay should be
observed. Also a few DPE Higgs events may be observed. The quality of a
diffractive event changes, however, at LHC energies. Besides the
production of a hard subsystem and one or two leading protons, the energy is
still enough for populating forward detector rapidity regions with particles.
As seen in Fig. \ref{fig-multiplicity}, the multiplicity of
particles is considerably
higher at the LHC, compared to the Tevatron. The requirement of
a ``clean'' diffractive Higgs event with a large rapidity gap in an observable
region cannot be achieved without paying the price of a lower cross section.
Requiring gaps instead of leading
protons gives a substantial reduction in the cross section, as seen in 
Table~\ref{tab-higgs}. Note that the high luminosity mode of LHC cannot be used, since the resulting pile-up of events would destroy the rapidity gaps.

The Monte Carlo model does not include any specific mechanism for the exclusive reaction $pp\to ppH$ and our simulations did not produce any such events. 

For comparison we have also investigated single diffractive Higgs production in
the pomeron model. This is based on the Regge framework with the exchange of a
Pomeron with vacuum quantum numbers \cite{GoulianosReview}, given by an
effective pomeron flux \cite{IS}. In case of a hard scattering process, which resolves an
underlying parton level process, a parton structure of the Pomeron may be
considered \cite{IS} and the data on diffractive deep inelastic scattering from HERA can be well described by fitting parton density functions in the Pomeron
\cite{HERA-pomeron}. Applying exactly the same model for $p\bar{p}$ gives,
however, diffractive hard scattering cross sections that are up to two orders
of magnitude larger than what is observed at the Tevatron. Although this can be cured by appropriately modified pomeron flux functions, it may indicate a deeper non-universality problem of the pomeron model \cite{SCI-TEV}. 

To get numerical estimates we use the pomeron model implemented in the \Pompyt \ Monte Carlo \cite{Pompyt}. The parton densities in the Pomeron are from a fit (parameterization I in \cite{GSpomeron}) to the diffractive structure function measured at HERA. The pomeron flux \cite{DLpomeron} has been renormalized \cite{fluxrenorm} so as to reproduce the observed relative rates of diffractive hard scattering processes both at HERA and the Tevatron. 

\begin{figure}
\begin{center}
\epsfig{width= 0.9\columnwidth,file=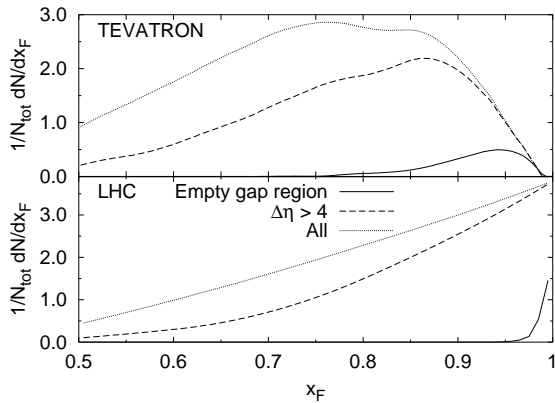}
\caption{Distribution in $x_F=p_{\parallel}/p_{{\rm max}}$ of leading protons in
single diffractive Higgs production in the pomeron model (\Pompyt{} Monte
Carlo) applied to Tevatron and LHC energies. Curves are for all such events
(dotted), events with no particles in $2.4<|\eta |<5.9$ (solid) and with a gap
of size at least four units in rapidity (dashed).}
\label{fig-pom-xf}
\end{center}
\end{figure}

The pomeron model is constructed to give a leading proton with a spectrum
essentially as $1/(1-x_F)$. It is developed for situations where $x_F\to 1 $
dominates and usually taken to be trustworthy only for $x_F> 0.9$. As shown in
Fig.~\ref{fig-pom-xf}, however, this distribution is strongly distorted in this
case due to the kinematical condition imposed by the Higgs mass. At the
Tevatron energy, the cross section is dominated by smaller $x_F$. This makes
the results of the model sensitive to a phase space region where the pomeron
model cannot be safely applied. In particular, the diffractive Higgs cross
section will depend on whether the usual requirement $x_F>0.9$ is applied or
not. The resulting cross section also depends strongly on what conditions for
diffraction are applied. The requirement at the Tevatron experiments of
no particles in the rapidity region $2.4<|\eta | < 5.9$ imposes a very strong
reduction. If the gap can be in a more forward rapidity region, based on
extended detector coverage, a much larger rate of
diffractive Higgs is obtained as illustrated in Fig.~\ref{fig-pom-xf}. Similar,
but not as strong effects are also present at LHC energies. 

In view of this, predictions for the diffractive Higgs cross section will be
somewhat uncertain in the pomeron model. To give some numbers, nevertheless, we use criterion (1) with a leading proton with $x_F>0.9$, but no specific gap requirement. This gives a cross section of $2.8\, \mbox{fb}$ for single diffractive Higgs production at the Tevatron and $410 \, \mbox{fb}$ at the LHC. This includes reduction factors of 5.2 and 9.2, respectively, from the pomeron flux renormalization \cite{fluxrenorm} making 
HERA and Tevatron data compatible but leaving an extrapolation uncertainty for
the LHC energy. 

In contrast to the pomeron model, the SCI and GAL models are constructed to
describe different final states through a general mechanism for soft color
exchanges giving a smooth transition between diffractive and non-diffractive
events. This implies a better stability with respect to variations of the
conditions used to define diffractive events. Moreover, the energy dependence of
SCI and GAL has proven successful. Data on various diffractive hard scattering
processes at HERA and Tevatron are well reproduced. The soft color exchange
models should, therefore, give more reliable predictions.

In conclusion, we have investigated the prospects for discovering the Higgs boson in diffractive events having a lower hadronic background activity that should simplify the reconstruction of the Higgs from its decay products. We find that the rate of diffractive Higgs events at the Tevatron will be too low to be useful. Therefore, the Higgs must here be searched for in normal events with their larger hadronic activity. At LHC diffractive events are not as clean as expected, since the large available energy produces an increased hadronic activity. Still, LHC should facilitate studies of Higgs in single diffraction and the observation of some DPE events with a Higgs boson.  


\end{document}